\documentclass[]{fairmeta}

\usepackage{graphicx}
\usepackage{textcomp}
\usepackage{xcolor}
    
\usepackage{url}
\usepackage{algorithm}
\usepackage{algpseudocode}
\usepackage{subcaption}
\newcommand{\newmaterial}[1]{{#1}}

\usepackage{xcolor}

\newcommand{\trace}{\textrm{tr}} 

\definecolor{mygray}{gray}{0.95}

\usepackage{amsmath,amsfonts,bm}

\def\eqref#1{equation~\ref{#1}}

\def\1{\bm{1}}

\def\vmu{{\bm{\mu}}}

\def\ve{{\bm{e}}}

\def\vp{{\bm{p}}}

\def\vr{{\bm{r}}}
\def\vs{{\bm{s}}}

\def\vx{{\bm{x}}}
\def\vy{{\bm{y}}}
\def\vz{{\bm{z}}}

\def\vsigma{{\bm{\sigma}}}
\def\vpsi{{\bm{\psi}}}

\def\eve{{e}}

\def\evs{{s}}

\def\evx{{x}}

\def\mE{{\bm{E}}}

\def\mW{{\bm{W}}}

\DeclareMathAlphabet{\mathsfit}{\encodingdefault}{\sfdefault}{m}{sl}
\SetMathAlphabet{\mathsfit}{bold}{\encodingdefault}{\sfdefault}{bx}{n}

\usepackage{tabularx,ragged2e,booktabs,caption}
\newcolumntype{C}[1]{>{\Centering}m{#1}}

\newcolumntype{Z}[1]{>{\Left}m{#1}}

\title{Beyond Self-Consistency: Loss-Balanced Perturbation-Based Regularization Improves Industrial-Scale Ads Ranking}

\author[1]{Ilqar Ramazanli\textsuperscript{*}, Hamid Eghbalzadeh\textsuperscript{*}, Xiaoyi Liu, Yang Wang, Jiaxiang Fu, Kaushik Rangadurai, Sem Park, Bo Long, Xue Feng}

\affiliation[1]{AI at Meta}
\contribution[*]{Equal Contribution}

\abstract{Perturbation-based regularization techniques address many challenges in industrial-scale large models, particularly with sparse labels, and emphasize consistency and invariance for perturbation in model predictions.
One of the popular regularization techniques has been various forms of self-consistency, which involve making small modifications to input data while preserving contextual information and enforcing similar predictions through auxiliary loss functions.
In this work, we explore the first successful application of perturbation-based regularization algorithms in large-scale ads ranking models, and further propose a novel regularization algorithm, namely, Loss-Balanced Small Perturbation Regularization (LSPR) that can be used in potentially any deep learning model.
We have successfully demonstrate that both Self-Consistency Regularization approaches (SCR) and LSPR are scalable and can improve ads delivery systems. 
By conducting industrial-scale experiments, and numerical analysis, we additionally show that our proposed LSPR, performs consistently better compared to SCR, across various groups and signal availability setups.
Finally, we report a successful application of the proposed LSPR in a billion-scale industrial ranking system, which to the best of our knowledge, is the first of its kind, and it is specially designed to address the various scalability challenges (e.g, various surfaces, geological locations, clients and so on) as we will mention in this paper. }

\date{2024-12-12}
\correspondence{Xue Feng at \email{xfeng@meta.com}}

\begin{document}

\maketitle

\section{Introduction} 
\label{sec:intro}
In the fast-paced and dynamic world of online advertising, the task of advertisements (ads) ranking helps businesses with their target audiences. 
The primary goal of ads ranking is to determine which ads are displayed to users via machine learning techniques, ensuring that the most relevant ones appears prominently. 
This process directly influences user engagement and click-through rates \cite{introduction_3, introduction_4}. 
\begin{figure}[h]
  \centering
  \includegraphics[width=6.5cm, height = 5.6cm]{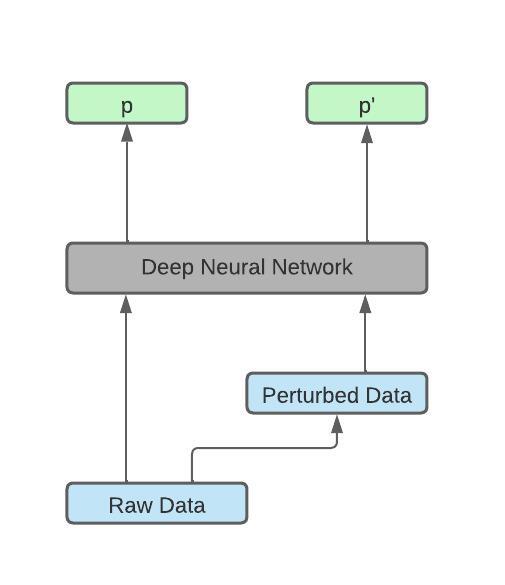}
  
  \caption{A General Perturbation Based Regularization Framework}
\label{fig:perturbation_framework}
\end{figure}

Ads ranking at an industry scale is often achieved through a multi-stage approach, encompassing retrieval, pre-ranking (or early-stage ranking), and final-stage ranking, which nowadays are mostly powered by large-scale neural networks \cite{introduction_1, introduction_2}.
This efficient multi-stage system strikes a balance between computational costs and recommendation quality \cite{int_ctr, int_ctr_2, naumov2019deep}. 

In recent years, the impact of deep learning, and notably its success in domains such as computer vision and natural language processing \cite{int_deep_learninig, int_nlp}, has been extended to recommendation systems.
Part of this success is due to the use of optimization objectives that can model user engagement via leveraging for deep neural networks, which as a result has motivated the migration of many significant industrial recommendation models to deep neural network architectures \cite{int_ctr_2, int_coll_dl}, illustrating its profound role in shaping the future of recommendation systems.

Self-supervised learning (SSL) stands out as a powerful technique with significant benefits for various facets of deep learning model development. 
At its essence, SSL is crafted to aid models in capturing intricate information that may prove challenging to extract directly from raw data, due to its reliance not only on labeled data which are often limited in amount, but also on unlabeled data which is more widely available. 
This capability becomes particularly pronounced when applied to large models facing constraints in accessing labeled data. 
Within the realm of SSL algorithms, the perturbation-based regularization technique emerges as a noteworthy one that is used jointly with various SSL techniques. 
This paper delves into an exploration of such regularization methods, shedding light on their roles and impacts within the broader domain of self-supervised learning for ranking models.

Various studies in the literature have shown the benefit for the use of simple input perturbations in regularizing model's generalization and robustness, which is dubbed in the literature as perturbation-based regularization (see Figure~\ref{fig:perturbation_framework}).
For instance, it has been shown that perturbing inputs with noise, regularizes the models towards more robustness and better generalization capabilities~\cite{pmlr-v139-dhifallah21a,pmlr-v206-orvieto23a, hua2021noise, NIPS2013_38db3aed}.
More concretely, two kinds of input perturbations have been identified to be effective in terms of model's generalization: 1) noise injection~\cite{pmlr-v139-dhifallah21a,pmlr-v206-orvieto23a, hua2021noise}, and 2) feature dropout~\cite{tamkin2022feature,NIPS2013_38db3aed,JMLR:v15:srivastava14a}.
Such regularizations have been proven to play an important role in preventing learning suppression, for instance, via leveraging techniques such as Self-Consistency Regularization e.g, in \cite{sinha2021consistency}, which evidently reports the distance between semantically similar points has undergone a significant reduction, showcasing the substantial impact of this regularization technique which leads to its popularity in the literature~\cite{ko2022self,tan2022hyperspherical,sinha2021consistency,wang2021deep,englesson2021consistency,kim2021selfmatch,kim2022conmatch}.

In this work, we take a broader look at perturbation based regularization approaches for industrial-scale applications in ads ranking, and share our findings on achieving better generalization via self-supervised learning, applicability for industrial usecases, and integration into complex industrial systems. 
More concretely, we present the first instance of its kind for a successful integration of perturbation based regularization into industrial-scale recommendation systems.
Additionally, we present \textbf{L}oss-Balanced  \textbf{S}mall- \textbf{P}erturbation  \textbf{R}egularization (\textbf{LSPR}), a novel perturbation-based regularization method that as we show, can improve the performance of industrial-scale ads ranking systems, while being simpler than its counterparts, hence, assist in scaling.
In summary, the main contributions of our work are as follows:

\noindent \textbf{Regularization Techniques for Ads Ranking at Scale}: 
We share our findings on regularization techniques that are applicable in industrial settings for ads ranking. 
These encompass improvements in offline metrics, and as we report, in several experiments we have obtained 0.1\% - 0.3\% relative Normalized Entropy (NE) offline gains by applying perturbation based regularizations.

\noindent \textbf{Loss-Balanced Small Perturbation Regularization (LSPR)}:
We propose LSPR: instead of adding an additional auxiliary loss function (e.g, often an MSE term) to alleviate the difference in predictions (e.g, as in Self-Consistency Regularization (SCR) ), we create new samples by perturbing datapoints with noise that are scaled by a small weight, and include them in the training data, but additionally weight them down in the the loss term calculation (see Figure~\ref{fig:lspr_diagram}).
Our numerical analysis (see Section~\ref{subsec:numerical_analysis}) shows LSPR achieves a better alignment with the optimal model parameters, and achieves lower errors in the model's weight space, compared to SCR.
Furthermore, we empirically verified (see Section~\ref{subsec:real_data}) that this technique performs better in large-scale industrial systems.
By applying this technique to several prediction models, we were able to achieve a 0.1\%-0.2\% relative NE gain.
We have additionally evaluated our approach in a set of online experiments, and have observed these offline performance improvements are also reflected in the online experimentation, which highlights our technique's effectiveness in online scenarios.

\noindent \textbf{Integration of Perturbation Based Regularization to Complex Industrial Systems}: 
To the best of our knowledge, this work on perturbation based regularization has been the first of its kind, to be integrated in industrial-scale recommendation systems for computing click-through/conversion rate prediction.
The process of incorporating data augmentation and self-supervised learning into complex architectures in large-scale industrial ads ranking and recommendation comes with its own set of challenges. 
Therefore, we provide system descriptions for large scale recommendation system, and how to navigate through its challenges, to adopt perturbation based regularization techniques optimally.
We further offer comprehensive design descriptions that encompass the data augmentation strategies and regularization algorithms we have experimented with, and  present the results we have achieved through these integrations (see Sections~\ref{subsec:numerical_setup} and \ref{sec:modeling}).

The remainder of paper is as follows. In Section~\ref{sec:related_work} we provide a literature review of the related topics.
The preliminaries are provided in Section~\ref{subsec:numerical_setup}.
We detail our modeling in Section~\ref{sec:modeling}.
Section~\ref{sec:experiments} will describe our experiment setup for numerical analysis and real data, and present their results. Finally, Section~\ref{sec:conclusion_related} will conclude the paper and provide insights on our future directions.

\section{Related Work}
\label{sec:related_work}
\subsection{Perturbation based self-Supervised learning} Perturbation based self-supervised learning has showcased its effectiveness in numerous applications.
For instance, Chen et, al. \cite{chen2020simple} introduced SimCLR, a Contrastive Learning approach, demonstrating that after representation learning with SimCLR, only a minimal 1\% of labeled data suffices to attain the same top-5 accuracy as AlexNet. Building on top of this work, Zbontar et, al. \cite{zbontar2021barlow} introduced Barlow Twins, which through the correlation of augmented and original data representations, achieved significant performance gains in computer vision problems. SSL has also made substantial contributions to the field of Natural Language Processing (NLP).  For instance, Gao et, al. \cite{gao2021simcse} introduced SimCSE, and Chuang et, al. \cite{chuang2022diffcse} introduced DiffCSE, both of which leveraged contrastive learning methods on improving sentence embeddings.

\subsection{Self-Supervised Learning for Recommendation Systems} With the substantial influence of perturbation based self supervised learning in the fields such as natural language processing and computer vision, researchers have extended their exploration to recommendation systems. One example of such kind of efforts is Wang et, al. \cite{contrastive_for_ctr} which focuses on enhancing Click-Through Rate (CTR) and Conversion Rate (CVR) estimation by applying Contrastive Learning techniques at the embedding level. This approach emphasizes the importance of post-embedding level operations and highlights the potential of self-supervised techniques for advancing ad ranking, offering valuable insights for large-scale ad recommendation systems.

In \cite{augmentation2}, researchers have made substantial contributions to the field of large-scale recommendation systems with a focus on perturbation based self-supervised learning. Their work introduces a two-stage perturbation approach at the embedding level, complemented by the application of contrastive learning to the predictions generated in each of these stages. Moreover, the paper introduces an inventive feature masking technique named Correlated Feature Masking. The combination of Correlated Feature Masking and Contrastive Learning yields exceptional performance in the desired metrics. These innovations, including the two-stage perturbation approach and Correlated Feature Masking, mark significant advancements in the domain of self-supervised learning for recommendation systems.

\cite{introduction_4} has harnessed the power of Self-Supervised Learning techniques in daily user interactions. Their work showcases that Self-Supervised Learning, combined with pre-training and fine-tuning, has led to impressive enhancements in Click-Through Rate (CTR) and Conversion Rate (CVR) tasks, yielding substantial improvements ranging from 6\% to 9\%. 

The strength of Self-Supervised Learning in recommendation systems has been comprehensively examined in the survey paper authored by Huang et al. \cite{self_rec_survey}. This survey provides an in-depth analysis of various Self-Supervised Learning methodologies, including Contrastive \cite{contrastive_survey}, Generative \cite{generative_survey}, Predictive, and Hybrid Methods. These techniques are thoroughly explored for their applicability in recommendation systems, offering valuable insights into the advancements of self-supervised approaches for this domain.

\subsection{Self-Consistency Regularization (SCR)}
Self-Consistency Regularization, engineered to ensure semantic similarity within the latent space for objects that share common semantics, as detailed in the research by Sinha et al. \cite{sinha2021consistency}, has a well-documented track record of efficacy. 
Previous studies consistently attest to the capability of this technique in fostering proximity of representations for semantically related objects in the latent space.
In the literature, one of the aspects that has been attributed to the success of consistency regularization and contrastive learning~\cite{zhang2022rethinking} has been identified as the use of Data Augmentation.

\subsection{Data Augmentation}
Data augmentation stands as a fundamental component in many self-supervised learning algorithms. While deep neural networks excel in various challenges in learning from data, they are particularly sensitive to data volume \cite{augmentation1} and often struggle to grasp the underlying data distribution. Given the scale of these models, insufficient data can lead to highly variable predictions in diverse settings.
Data augmentation can incorporate strong priors from data or domain knowledge into models~\cite{eghbalzadeh2024rethinking}, and further be used to regularize models towards better robustness and generalization~\cite{zhang2017mixup,yun2019cutmix}. However, most of the focus in such approaches have been on structured data such as images, audio, etc; and it has been shown that such domain-specific augmentations should be used in new domains with caution~\cite{eghbalzadeh2024rethinking}.

\section{Preliminaries}
\label{subsec:numerical_setup}
Click-Through Rate (CTR) prediction aims to estimate the probability of the user clicking a candidate ad after having an impression in the ranking stage. Similarly, Conversion Rate (CVR) prediction estimates how likely the user will convert the candidate ad after having a click. Our perturbation-based regularization techniques can be applied to both CTR and CVR predictions with similar set-ups, therefore, we use the CTR prediction as an example to introduce the basic preliminaries, and the intrinsic differences between CTR and CVR modeling (e.g., delayed feedback for ad conversions) is beyond the scope of the discussions in this paper.

For the CTR prediction task, let a training dataset with $\mathit{N}$ examples be defined as $\left\{\vx_n, {\vy}_n\right\}_{n=1}^{N}$, where a random variable $\vx_n$ represents the feature space of the $n$-th training example, and a random variable $\vy_n\in \{0, 1\}$ represents the binary label indicating whether the user has clicked the candidate ad or not. The feature space can consist of the following types:
\begin{itemize}
    \item \textbf{Dense features} are single-digit float values (e.g., counts and stats (mean, percentiles, variances) of user/ad behaviors and profiles), and the total number of such features could be in the scale of thousands. 
    We initially apply a pre-processing procedure on each of them, and then concatenate them together to form a single high-dimensional float vector, so as to interact with other features later;\\
    
    \item \textbf{Embedding features} are high-dimensional float vectors which are usually generated from pre-trained manners (e.g., user and ad embeddings from graph learning algorithms). We will denote embedding features as $\ve_i$\\ 
    
    \item \textbf{Sparse features} are high-dimensional integer vectors, that represent concepts such as user-item interactions, where both number of items and users are large. Sparse features can often be represented via a much lower dimensional vector $\vs_i$ via various techniques, e.g, the use of an embedding matrix, or affinity scores for reweighing.
    
\end{itemize}

After processing each feature to generate the corresponding representation, the feature interaction layer is applied on top to learn their interactions with arbitrary orders (e.g., DHEN \cite{DHEN}, DCN \cite{DCN}, Transformer \cite{Transformer}), and generate a final representation $\vr_n$. Afterwards, the prediction layer produces the prediction probability $\hat{\vy}_n \in [0, 1]$ based on $\vr_n$, and the commonly adopted loss function is calculated as  

\begin{align}
\label{eq:bce}
   \mathcal{L_{\text{supervised}}}(\vy_i,\hat{\vy}_i) = -\frac{1}{N}\sum_{n = 1}^{N} \vy_n \log (\hat{\vy}_n) + (1 - \vy_n) \log (1 - \hat{\vy}_n).
\end{align}

\section{Methodology}
\label{sec:modeling}
In this section, we delve into the core regularization techniques we have applied to an industrial-scale ads recommendation system. 
We start by discussing our data augmentation strategies which are an important part of Perturbation-Based Regularization, and further detail Self-Consistency regularization methods. 
We then discuss regularization techniques that promote perturbation invariance beyond Self-Consistency Regularization.
Finally, we discuss the integration of these techniques into different phases of industrial-scale models - such as Retrieval, Early and Final Stage Ranking.

\subsection{Data Augmentation} 

Data augmentation has played a role in our perturbation-based regularization algorithms. 
It's essential to underline that recommender systems, as mentioned in \cite{int_ctr} and \cite{augmentation2}, are significantly influenced by both sparse and dense features. 
Therefore, a robust augmentation strategy that caters to both types of features, improving the effectiveness of our regularization methods in various recommendation scenarios.

\textbf{Dropout}  
was initially proposed as a regularization method that enabled deep learning models to generalize, and is known as one of the stepping stones of deep learning~\cite{srivastava2014dropout}.
It further emerged as a vital data augmentation strategy tailored for sparse features, leveraging insights from its prior applications in natural language processing \cite{gao2021simcse} and recommender systems \cite{augmentation2}. In this context, the core concept involves creating a subset of existing sparse features as augmented copies of the original sparse feature set.
For instance, consider a datapoint with embedding features $\ve$ and sparse features $\vs$:
\begin{align*}
    (\evs_1,\evs_2,\evs_3,\evs_4,\evs_5,\evs_6) \rightarrow  (\eve_1, \eve_2, e_3), 
    (\evs_1,\evs_2,0,0,\evs_5, \evs_6) 
\end{align*}

The extent of dropout perturbation varies depending on the problem setting, with the option to employ either a strong or weak dropout. When integrated correctly with Self-Supervised Learning (SSL) techniques, dropout has exhibited substantial performance improvements in large-scale item recommendations \cite{augmentation2}, emphasizing its pivotal role in enhancing recommendation systems.

\textbf{Gaussian Noise Injection} serves as a technique for augmenting dense features within our framework. The concept is elegantly simple, involving the generation of a random vector $\vpsi_i$ from a Gaussian distribution, denoted as $\vpsi_i\sim\mathcal{N}(\vmu, \vsigma)$. 
For instance, in a 3-dimensional float vector, represented as $\vx = (\evx_1, \evx_2, \evx_3)$, augmentation with Gaussian Noise can be described as follows:
\begin{align*}
(\eve_1, \eve_2, \eve_3) \rightarrow (\eve_1 + \psi_1, \eve_2 + \psi_2, \eve_3 + \psi_3)
\end{align*}

This augmentation introduces controlled randomness to the features, contributing to the model's robustness and diversity of the data.

\subsection{Self-Consistency Regularization (SCR)}

\begin{figure}[ht!]
  \centering
  \includegraphics[width=7.5cm, height = 7.8cm]{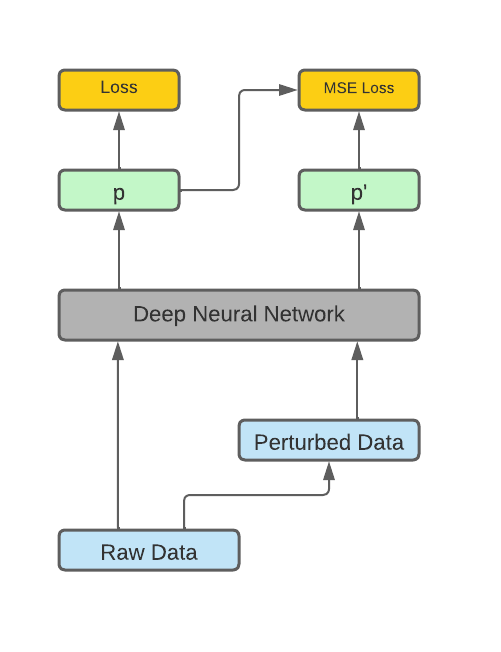}
  \caption{Self Consistency Regularization (SCR)}
\label{fig:consistency}
\end{figure}

Self-Consistency Regularization is an algorithm that enforces small modifications in the data still preserve the similar prediction value.
The algorithm, is especially important when the model is too large, and data-set is not large enough to serve to the model's capacity.
The algorithm introduces an auxiliary loss term, that penalizes the disparity between the outcomes of the perturbed and the original data point, effectively promoting consistency in the latent space representation (see Figure~\ref{fig:consistency}).

The concept underlying SCR is as straightforward as depicted in Figure~\ref{fig:consistency}.
As can be seen, perturbed data along with the original data is fed to the model, and an additional regularization loss is used to minimize the model's output differences between original and perturbed data.
In this approach, we incorporate Mean Squared Error (MSE) loss term as the regularizer alongside the supervised loss term.
\begin{align*}
    \mathcal{L_{\text{consistency}}}(\vy_i,\hat{\vy}_i,\vp_i,\vp'_i)  = \mathcal{L_{\text{supervised}}}(\vy_i,\hat{\vy}_i) + \lambda \mathcal{L_{\text{MSE}}}(\vp_i,\vp'_i)
\end{align*}
where $\mathcal{L_{\text{supervised}}}$ is as defined in Eq.~\ref{eq:bce} and $\mathcal{L_{\text{MSE}}}$ is defined as
\begin{align*}
\mathcal{L_{\text{MSE}}}(\vp_i,\vp'_i)&= \frac{1}{N} \sum_{i=1}^{N}(\vp_i- \vp'_i)^2\\
\end{align*}
and $\vp_i$ represent some hidden representation of the deep neural network for some input $\vx_i$, while ${\vp'}_i$ denotes this representation for the perturbed input ${\vx'}_i$.

\subsection{Loss-Balanced Small Perturbation Regularization (LSPR)}
\begin{algorithm}
\caption{Loss-Balanced Small Perturbation Regularization}
\label{alg:lspr}
\begin{algorithmic}
\For{ batch from data}
    \State 1. Sample data
    \State 2. Sample small noise
    \State 3. Create perturbed data by adding noise to data
    \State 4. Calculate $\mathcal{L_{\text{supervised}}}$ on sampled data
    \State 5. Calculate $\mathcal{L_{\text{supervised}}}$ on perturbed sampled data
    \State 6. Balance losses by calculating $\mathcal{L_{\text{LSPR}}}$ from Eq.~\ref{eq:spr}
    \State 7. Update model parameters
\EndFor

\end{algorithmic}
\end{algorithm}
Despite its simplicity and generality, training models with noise has been known to improve  generalization of models~\cite{bishop1995training}.
In this section, we study a variation of Perturbation-Based Regularization, namely, Loss-Balanced Small Perturbation Regularization (LSPR). 
In this approach, perturbed points are treated as original points but with smaller weights in the loss.
Moreover, we expect these datapoints that contain small perturbations to have the same label as the original data points. Therefore, we name this algorithm Loss-Balanced Small Perturbation Regularization (LSPR).
In contrast to data augmentation that treats both augmented (or perturbed) data and original data equal in the loss calculation, LSPR reduces the weights of perturbed data in the calculation of the loss, hence, is less disruptive to learning dynamics.
\newmaterial{
Furthermore, we report a successful deployment of LSPR in a billion-scale industrial ranking system.
To the best of our knowledge, LSPR is the first of its kind, and it is specially designed to address the various scalability challenges. Not only does the system need to cater to billions of users, but also serve various surfaces (e.g, client-facing apps and product platforms), global geological locations, various clients (e.g, web, mobile app), and various conversion events (e.g. clicks, purchases) which means the system consists of hundreds of models up and running at any given time.}

The LSPR algorithm is depicted in Algorithm~\ref{alg:lspr}. 
\newmaterial{As can be seen, LSPR constructs perturbations to create perturbed examples, then uses those perturbation to calculate a regularization loss, which is combined with the main objective and is balanced accordingly.
In constructing the perturbations, LSPR ensures that perturbation and data are both of the same class of distributions. For instance, if data is categorical, the perturbations will also be of a categorical distribution.}

In this paper, we have treated any perturbed data point with a uniform weight. Here, we scale samples uniformly with a scale parameter $\lambda<1$. 
However, exploring perturbation-dependent weights is a worthwhile follow-up.
The formula for LSPR regularization is as follows:
\begin{align}
\label{eq:spr}
    \mathcal{L_{\text{LSPR}}}(\vy_i,\hat{\vy}_i, \hat{\vy}'_i)   = \mathcal{L_{\text{supervised}}}(\vy_i,\hat{\vy}_i) + \lambda \mathcal{L_{\text{supervised}}}(\vy_i,\hat{\vy}'_i) 
\end{align}

where $\hat{\vy}'_i$ is the model's prediction on the perturbed input $\vx'_i$
traditional supervised loss function.
The schematic representation of this regularization technique is outlined below in the Figure~\ref{fig:lspr_diagram}:
\begin{figure}[h]
  \centering
  \includegraphics[width=7.5cm, height = 8.1cm]{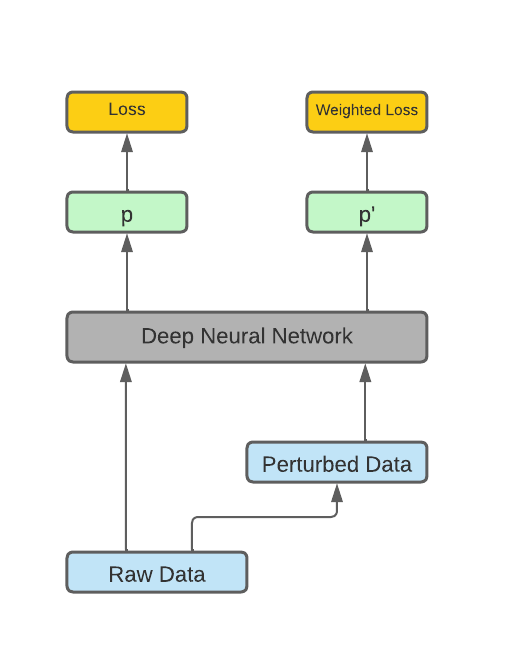}
  \caption{Loss-Balanced Small Perturbation Regularization. (LSPR)}
\label{fig:lspr_diagram}
\end{figure}
Unlike Self-Consistency Regularization, Small Perturbation regularization does not prioritize minimizing the distance between original data points and perturbed points. 
Instead, it focuses on correct predictions for perturbed points.
Depending on the defined loss $\mathcal{L}$, this discrepancy can result in significant variance in the resulting parameters.
Hence, the batch size at each stage will be doubled in this case, while some of the points having smaller weights compared to others.

\subsection{LSPR's Hyperparameters}

\newmaterial{
LSPR is designed to be simple yet effective, with only three major hyperparameters, while providing significant values in performance and optimization. Here are its hyperparameters and our approach in hyperparameter tuning:
- dense feature perturb: We enforce perturbation to be of the same distribution as our dense features.
- sparse feature dropout: we apply a relatively small dropout rate to sparse features.
- loss weight: We start our hyperparameter search for the loss weight from a smaller scale relative to main objectives, and then to a more fine grained search
This simplicity allows for easier tuning and deployment in large-scale industrial settings while still delivering significant performance improvements.
We will add these details to the final version of our paper.
}

\section{Analysis and Experimentation }
\label{sec:experiments}
\newmaterial{
In this section, we leverage a well-known theoretical framework proposed in~\cite{werfel2003learning} to demonstrate how LSPR results in a better alignment of weights in the model optimization to portray a clear picture of the construct of a optimization problem in ranking, and how LSPR affects it.}
We start by formalization of our framework, as well as the integration of Perturbation-Based Regularizations, namely SCR and LSPR. 
In Section~\ref{subsec:numerical_analysis} we analyze how LSPR compares to SCR via controlled experimentations and analysis on linear models, investigating the learning dynamics with these regularization applied.
Further, in Section~\ref{subsec:real_data}, we report our empirical results on an in-house dataset that was used to evaluate the methodologies applied here. 
We tracked model accuracy using Normalized Entropy (NE) in offline experiments \cite{he2014practical}.
In experiments with real data, each datapoint exhibits a substantial volume of features, comprising thousands of dense features and hundreds of sparse features and we employ the Adagrad optimizer for optimization.
Ranking has been done through multiple stages during learning, which are described below in more details.
In this section, we report performance improvements via the presented regularization techniques on a multi-stage ranking system with 3 stages of retrieval, early stage ranking, and final-stage ranker.

\subsection{Numerical Analysis}
\label{subsec:numerical_analysis}

In this section, we provide a numerical analysis for the linear models trained with SGD 1) with Self-consistency Regularization (SCR), and 2) with Loss-Balanced Small Perturbation Regularization (LSPR).
We analyse the gradient update directions and the alignment with the optimal weight (See Section.~\ref{subsec:numerical_setup}) by calculating the cosine similarity in the model's weight space, comparing weights of different iterations to the optimal weight.
Our numerical analysis (see Figure~\ref{fig:numerical_analysis}) shows that: 
\begin{enumerate}
    \item compared to SCR, LSPR finds a better alignment with the optimal weight, while converging faster and achieving a lower error in the weight space. 
    \item we also show that balancing both amount of noise $\omega$ and loss $\lambda$ is crucial to the success of LSPR and SCR. As we show, smaller values for these weights are recommended for better convergence and performance.
\end{enumerate}

\subsubsection{Setup}
The goal in this section is to analyse how different perturbation-based regularizations, namely SCR and LSPR, impact learning and performance.
To this end, we simplify both LSPR and SCR frameworks to their core, and furthermore using linear models study their effects in learning dynamics and performance. We use 2-layer linear models which strike a good balance between model expressiveness and simplicity~\cite{werfel2003learning}. To this end, we define a ground-truth function with the weight $\mW^*$ that maps input data to their labels as follows:
\begin{align}
    \vy={\mW}^*\vx
\end{align}
where $\vx$ denotes an input feature and $\vy$ denotes the ground-truth output, and ${\mW}^*$ is a $L_y\times L_x$ matrix where $L_y$ and $L_x$  are input and output dimensionalities.

We now define the following linear model that we use to learn the input-output relationship by:
\begin{align}
    \vy={\mW_2}{\mW_1}\vx
\end{align}
where $\vx$ denotes an input feature and $\vy$ denotes the ground-truth output,  ${\mW_1}$ is a matrix of size $L_h\times L_x$ and  and $\mW_2$ is a matrix of size $L_y\times L_h$ and $L_h$ is the dimensionality of the intermediate representations.
To simulate the effect of regularization, we use Stochastic Gradient Descent (SGD) with an MSE error as follows:
\begin{align}
    \mathcal{L}(\vx,\vy)=\frac{1}{2}||\vy-\hat{\vy}||^2,
\end{align}
and $\hat{\vy}$ is the output of the linear model.
In order to study the learning dynamics, we denote the  weight error as:

\begin{align}
    \mE=\mW_2\mW_1-\mW^*
\end{align}
and further introduce:
\begin{align}
    \epsilon=\frac{1}{L_x L_y}\trace[\mE^T\mE], \gamma=\frac{{\mW_2\mW_1 \cdot \mW^*}}{{\|\mW_2\mW_1\| \|\mW^*\|}}
\end{align}
where $\epsilon$ represents the error in the weight space to the optimal weight, while $\gamma$ demonstrates weight alignment with the optimal weight $\mW^*$.
The SGD weight updates are as follows:
\begin{align}
    \delta\mW_1^{SGD}&=-\eta\frac{\partial \mathcal{L}(\vx,\vy)}{\partial \mW_1} \\
    &=-\eta\left(\mW_2^T(\mW_2\mW_1\vx-\vy)\right) \otimes \vx \\
    \delta\mW_2^{SGD}&=-\eta\frac{\partial \mathcal{L}(\vx,\vy)}{\partial \mW_2} \\
    &=-\eta(\mW_2\mW_1\vx-\vy) \otimes \mW_1\vx 
\end{align}
with $\otimes$ denoting the outer product.

In order to simulate LSPR, we sample noise $\vz\sim\mathcal{N}(0, I)$ and additionally add $\mathcal{L}(\vx+\omega\vz,\vy)$ where $\omega$ is a small weight for the perturbation $\vz$.
The final loss will be a balance of $\mathcal{L}(\vx,\vy)+\lambda\mathcal{L}(\omega\vz+\vx,\vy)$.
The LSPR weight updates are then defined as:
\begin{align}
\label{eq:lspr_updates}
    \delta\mW_1^{LSPR}&=-\eta\left(\mW_2^T(\mW_2\mW_1\vx-\vy)\right) \otimes \vx \nonumber\\
&-\lambda\eta\left(\mW_2^T(\mW_2\mW_1(\omega\sigma\vz+\vx)-\vy)\right) \otimes (\omega\sigma\vz+\vx)\\  
\delta\mW_2^{LSPR}&=-\eta(\mW_2\mW_1\vx-\vy) \otimes \mW_1\vx \nonumber\\
&-\lambda\eta(\mW_2\mW_1(\omega\sigma\vz+\vx)-\vy) \otimes \mW_1(\omega\sigma\vz+\vx)
\end{align}

To analyse the SCR method, we rely on the additional learning signal that pushes the output of a model on clean and noisy inputs closer together, namely, $\mathcal{L}(\vx+\omega\vz,\hat{\vy})$ where $\hat{\vy}=\mW_2\mW_1\vx$.
Consequently, the SCP weight updates are as follows:
\begin{align}
\label{eq:scr_updates}
    \delta\mW_1^{SCR}&=-\eta\left(\mW_2^T(\mW_2\mW_1\vx-\vy)\right) \otimes \vx \nonumber\\
&-\lambda\eta\left(\mW_2^T(\mW_2\mW_1(\omega\sigma\vz+\vx)-\mW_2\mW_1\vx)\right) \nonumber\\
&\otimes (\omega\sigma\vz+\vx)\\  
\delta\mW_2^{SCR}&=-\eta(\mW_2\mW_1\vx-\vy) \otimes \mW_1\vx \nonumber\\
&-\lambda\eta(\mW_2\mW_1(\omega\sigma\vz+\vx)-\mW_2\mW_1\vx) \nonumber\\
&\otimes \mW_1(\omega\sigma\vz+\vx)
\end{align}

We use the following parameters for our analysis: $\omega=\{0.1,0.9\}, \lambda=\{0.001,1\}, \eta=1.4, L_x=100,L_h=10^4,L_y=10$, and we perform the weight updates for the number of $100k$ times, and for every update we sample new data from the denoted distributions.

\subsubsection{Results}
\begin{figure*}[t]
    \centering
    \begin{subfigure}{0.45\textwidth}
        \centering
        \includegraphics[width=\textwidth]{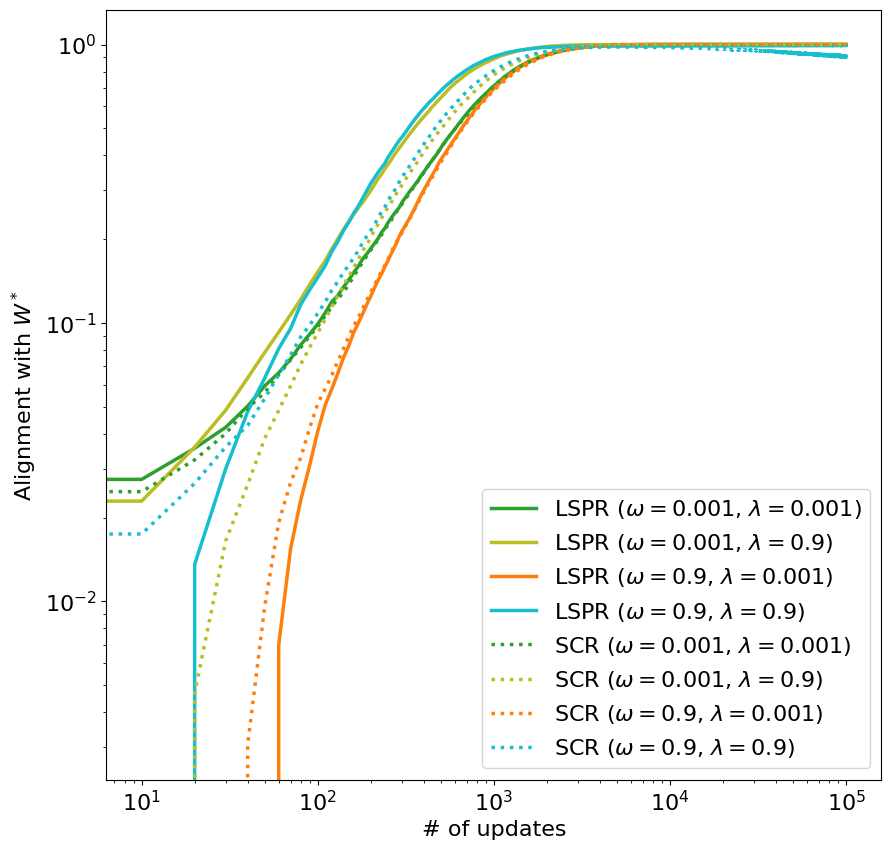}
        \caption{Alignment with the optimal weight $\mW^*$.}
        
        \label{fig:first_fig}
    \end{subfigure}
    \hfill
    \begin{subfigure}{0.45\textwidth}
        \centering
        \includegraphics[width=\textwidth]{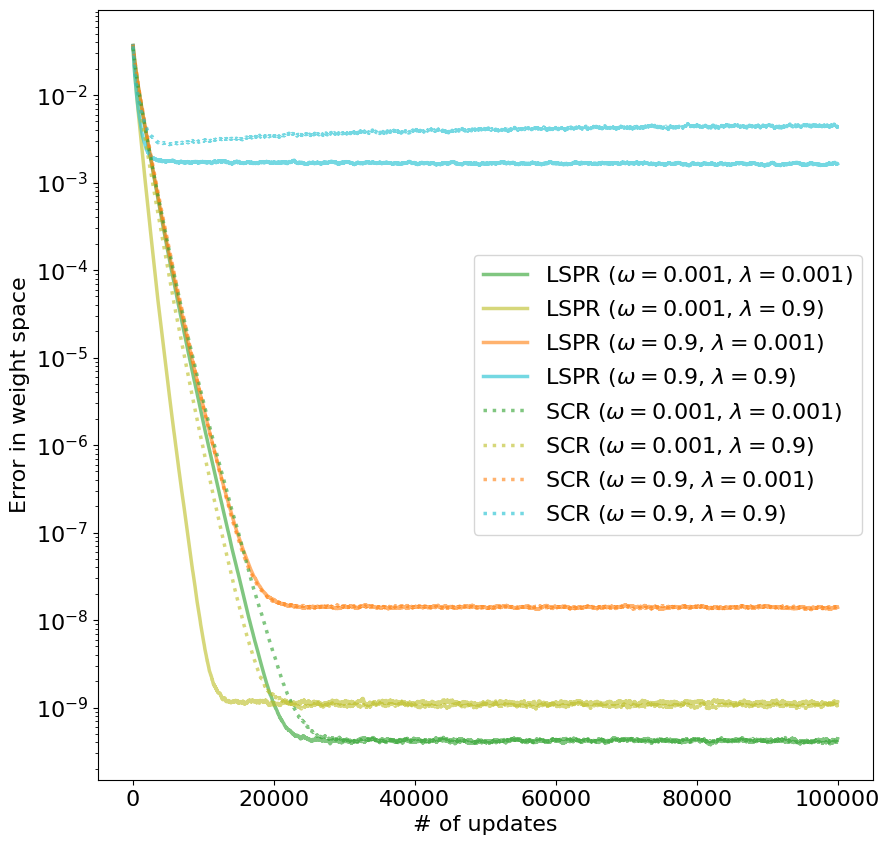}
        \caption{Error in the weight space.}
        \label{fig:second_fig}
    \end{subfigure}
    \caption{Numerical analysis comparing LSPR and SCR.
    $\omega$ denotes noise sample weight and $\lambda$ depicts loss weight.} 
\label{fig:numerical_analysis}
\end{figure*}

As can be seen in Figure~\ref{fig:numerical_analysis}, we can observe that for small perturbation weights $\omega$ and loss weights $\lambda$, LSPR tends to better find the optimal weight as can be seen by looking at the two presented plots.

\subsection{Experiments on Real Data}
\label{subsec:real_data}
\subsubsection{Self-Consistency Regularization (SCR)}

We experimented with perturbation-based consistency regularization on different stages of various prediction problems. We present these results in Table~\ref{tab:rel_ne}.
We observed a relative NE gain of approximately 0.1\%-0.3\%, depending on the prediction model tested in various offline experiments.
We will first present the results for the Retrieval stage from the offline experiments, followed by the experimental results for the Early and Final Stage ranker.

\noindent\textbf{Offline Retrieval Stage:} We have experimented with consistency regularization in two different models that predict conversion rate and click-through rate, respectively. 
We obtained the best results when regularizing both logit and representative of embedding with 0.15\%-0.2\% relative NE improvements.

\noindent\textbf{Offline Early Stage Ranking:} models are generally simpler ranking models compared to final stage models. Therefore, we applied regularization to the entire object and user embedding, resulting in a 0.3\% relative NE gain in various offline experiments.

\noindent\textbf{Offline Final Ranking Stage:} 
Models in this stage are generally much larger and complex compared to previous stages, as we are looking for more precise ranking of ads. 
We obtained the best results by regularizing both the logit and output of the embedding together.
Results from several experiments suggest an average 0.1\% relative NE gain, which has been further validated with online testing.

\begin{table}
  \caption{Relative NE gains for SCR across various stages.}
  \label{tab:rel_ne}
    \centering
  \begin{tabular}{cccc}
  \toprule
    Model & 33\% of data &  66\% of data & 100 \% data\\
    \midrule
    \texttt{Baseline} & 0 \% & 0 \% &  0 \% \\
    \textbf{Retrieval} & \bf{0.14 \%} & \bf{0.19} &  \bf{0.14 \%}\\
    \textbf{Early Stage} & \bf{0.28 \%} & \bf{0.3 \%} &  \bf{0.35 \%}\\
    \textbf{Final-stage Ranker} & \bf{0.1 \%} & \bf{0.08 \%} &  \bf{0.07 \%}\\
    \bottomrule
  \end{tabular}
\end{table}

\subsubsection{Loss-Balanced Small Perturbation Regularization (LSPR)}
We have explored LSPR primarily in the offline Final Ranker Stage, under various signal availability setups. We have observed that the technique has performed promisingly in various setups, ultimately leading to improved performance in each of these environments. These results are depicted in Table~\ref{tab:rel_ne2}.

\noindent\textbf{Offline Final Ranking Stage:} testing is very similar to Consistency Regularization testing; however, in the former, we perturbed the entire batch each time, leading to doubled batch size. 
In contrast, for Consistency Regularization, we only perturbed a small fraction of points in each batch.\\

\begin{table}
  \caption{Relative NE gains comparing SCR and LSPR on Final-stage Ranker.} 
  \label{tab:rel_ne2}
  \centering
  \begin{tabular}{cccc}
    \toprule
    Model & 33\% of data &  66\% of data & 100 \% data\\
    \midrule
   \texttt{Baseline} & 0 \% & 0 \% &  0 \% \\
    \textbf{SCR} & \bf{0.1\ \% } & \bf{0.08 \%} &  \bf{0.07 \%}\\
    \textbf{LSPR} & \bf{0.13 \% } & \bf{0.11 \%} &  \bf{0.1 \%}\\
    \bottomrule
  \end{tabular}
\end{table}

\subsubsection{Online Experiments}

We additionally have conducted online experimentation for a prediction model after testing it in offline setup.
The online experimentation is different than offline one in the nature that, it runs in continuous training and inferring routine, compared to full training and inferring mode.
These online experiments on various data from different parts of the data stream, using both noisy and clean labels, have demonstrated a similar trend to the offline experiments we reported in the previous sections.
\newmaterial{
Our results indicate that LSPR has achieved a 0.1\% to 0.2\% relative improvement in online top-line metrics, consistently across multiple launches. Note that the magnitude of the impact is significant at the level of a billion-scale industrial production ads ranking system, which serves billions of users across various surfaces , across global geological locations, and across various clients. 
}

\subsection{Baselines}

\newmaterial{
The experiment comparisons in this manuscript are all compared against the latest production models in a multi-billion-scale industrial ads ranking system, prior to the adoption of LSPR. 
Our criteria for selecting baselines was to identify models that 1) have been proven to operate effectively at the industry scale; 2) represent the  state-of-the-art ads ranking product models  in the industry.
We consider these production-level recommendation models to be among the state-of-the-art baselines that meet the above criterion. 
}

\section{Conclusion and Future Work}
\label{sec:conclusion_related}

Our study has explored the application of perturbation based regularization algorithms in an Industrial-Scale Recommendation Systems. 
To this end, we have made significant contributions: firstly, to the best of our knowledge, we showed for the first time that Perturbation Based Regularization techniques can bring meaningful improvements to Industrial-Scale Recommendation Systems.
Secondly, we introduced a novel regularization technique - LSPR,a general method that is applicable in many Deep Learning setups.
\newmaterial{In summary, LSPR has been launched to major industrial-scale ads recommendation models across different ranking stages and traffic. This indicates that it can be generalized to diverse user demographics and content types, considering the scale and reach of the deployed ads platform.}
Our future research endeavors are poised to focus on other variations of the use of unlabeled data, tailored for Large Scale Recommendation Systems, pushing on both theoretical understanding, as well as industrial-scalability.
These next steps represent our commitment to pushing the boundaries of recommendation systems, with a keen focus on understanding and optimizing ad recommendations to better serve both users and businesses.

\clearpage
\newpage
\bibliographystyle{assets/plainnat}
\bibliography{refs}

\end{document}